# Nonequilibrium ac Josephson effect in mesoscopic Nb-InAs-Nb junctions

K.W. Lehnert,[1,2] N. Argaman,[2,3] H.-R. Blank,[2,4] K.C. Wong,[2,4] S.J. Allen,[1,2] E.L. Hu,[2,4] H. Kroemer[2,4]
[1]*Department of Physics and Quantum Institute*
[2]*QUEST, the NSF Center for Quantized Electronic Structures*
[3]*Institute for Theoretical Physics*
[4]*Department of Electrical and Computer Engineering, University of California,
Santa Barbara, CA 93106*
(Received July 27, 1998)

Microwave irradiation of Nb-InAs-Nb junctions reveals frequency-doubled Josephson currents which persist to high temperatures, in the absence of a critical current. A nonequilibrium dynamical model, based on time-dependent Andreev bound states, successfully accounts for the resulting half-integer Shapiro step and an enhancement in the conductance near zero bias.

Superconducting weak links and proximity effects have recently attracted much renewed interest due to the development of fabrication techniques for junctions of mesoscopic dimensions. For example, Courtois *et al* [1] have demonstrated the coexistence of different temperature dependences of different proximity effects, and Scheer *et al* [2] have obtained detailed agreement between the theoretical and experimental dc characteristics of atomic-sized superconducting constrictions. In this letter, we extend this trend to dynamical nonequilibrium (NEQ) supercurrents [3-4].

We have found that in mesoscopic superconductor-normal metal-superconductor (SNS) junctions where the normal metal is a two-dimensional electron gas (2DEG) formed in an Indium Arsenide (InAs) quantum well, ac supercurrents flow in the absence of any dc Josephson effect, and at twice the usual frequency. This remarkable behavior is a dynamical NEQ effect in which the time dependence of the difference in the phases $\phi$ of the two superconducting order parameters drives the electronic states in the normal metal out of equilibrium, revealing the phase coherent transport hidden by thermal smearing. The signature of this effect is a half-integer Shapiro step which persists to temperatures well above the temperature at which both the critical current $I_c$ and the integer Shapiro step disappear. These results agree qualitatively with theories [5-6] of dynamical NEQ. Additionally, a quantitative NEQ model demonstrates that the half-integer step and a conductance enhancement around zero voltage bias [5-7] are different aspects of the same phenomenon.

The devices we study consist of two superconducting Nb electrodes resting directly on top of the InAs layer and coupled by a full AlSb-InAs-AlSb quantum well. Details of the quantum wells are published elsewhere [8]. The *unprocessed* 2DEG had an elastic scattering length $l=4$ μm and Fermi velocity $v_F=10^6$ m/s. InAs does not form a Schottky barrier with most metals, enhancing the probability of Andreev reflection [9]. All of the data shown come from a single device with a width perpendicular to the current flow of 100 μm, and electrode separation $L=1.2$ μm, but other coprocessed devices show the same behavior [10].

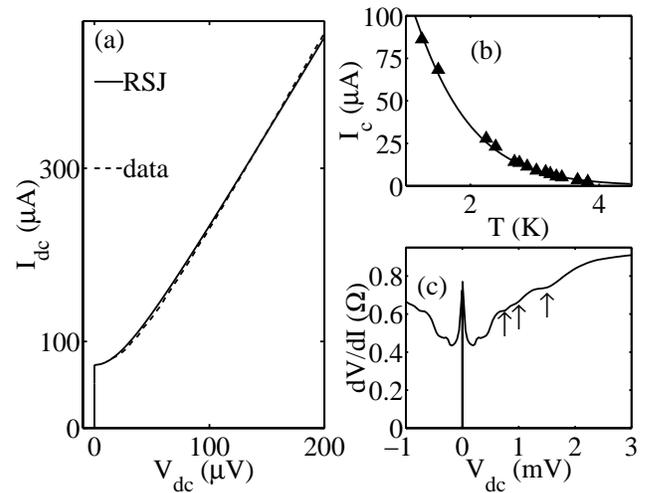

FIG. 1. (a) CVC at 1.4 K and RSJ model fit. $R_N=0.43$ Ω and $I_c=75$ μA. (b) $I_c$ versus $T$, ▲. The line is a fit to $I_c(T) \propto T^{1/2}\exp(-2\pi kT/E_c)$, $E_c/k=(4.1\pm0.1)$ K. (c) $dV/dI$ versus $V_{dc}$ at 1.4 K. Arrows indicate SGS at $2\Delta/en$ for $n=2,3,4$ and $2\Delta/e=3$ mV, where $2\Delta$ is the full BCS gap in Nb.

The current-voltage characteristic (CVC) of this single junction shows $I_c=75$ μA at 1.4 K, but $I_c(T)$ is exponentially suppressed with increasing $T$ (Fig. 1). A fit to the clean limit [11] form, $I_c(T) \propto T^{1/2}\exp(-2\pi kT/E_c)$ where $\hbar v_F/2\pi kT<<l,L$ gives the Thouless energy as $E_c=\hbar/\tau_d=350$ μeV where $\tau_d=1.8$ ps is the transit time of an electron across the junction. Calculating $\tau_d v_F/L=1.5$ indicates that in contrast to previous study of dynamical NEQ in dirty, diffusive systems [3-4] these devices are clean and quasiballistic, $L\approx l$ [12]. The differential resistance $dV/dI$ shows subgap structure (SGS) [Fig. 1(c)], the hallmark of multiple Andreev reflections, and is consistent with a superconductor-semiconductor interface that is very transparent to electrons [13]. From these dc data, the resistively shunted junction (RSJ) model [14] would *appear* to provide an adequate description of the junction dynamics in a range of dc voltage bias $V_{dc}<200$ μV [Fig. 1(a)], cf.[3].

We probe ac currents at the *n*th harmonic of the Josephson frequency, $2neV_{dc}/h$, by irradiating the junction with microwaves at a nearby frequency ν and looking for phase-locking effects, called fractional Shapiro steps (integer steps occur when ν or a harmonic of ν, resonates with $2eV_{dc}/h$). The measured $V_{dc}$, for a given dc current bias $I_{dc}$, changes by an amount $\Delta V$ when a small microwave current bias $I_{ac}$ is applied. The maximum value of $\Delta V$, is $R_d \Lambda I_{ac}$ in the absence of noise, where $R_d$ is the differential resistance at $V_{dc}=h\nu/2en$, the location of the Shapiro step, and $\Lambda$ is the dimensionless magnitude of the step. We focus on the regime of weak microwave current drive, where the steps are broadened by noise. In this limit,

$$\Delta V = -\frac{\Lambda^2}{2} R_d \left( \frac{I_{dc} - I_0}{(I_{dc}-I_0)^2 + (I_\Gamma)^2} \right) I_{ac}^2 \quad (1)$$

where $I_\Gamma = 2\pi ek T_N (R_d/R_N)/\hbar$ defines the noise temperature $T_N$, and $I_0$ is the value of $I_{dc}$ at the center of the Shapiro step. In the RSJ model $\Lambda = I_c/2I_0$, but Eq. 1 with different expressions for $\Lambda$ is expected to hold in other models as well [15].

The response of the junction to pulsed microwave irradiation, measured with lock-in techniques, is shown in Figs. 2 and 3. The sample was enclosed in a superconducting can pierced by measurement wires and a coaxial cable with an open termination inside the can. Measurable coupling of microwaves to the device occurred only near resonant frequencies of the resulting microwave cavity. We assume ac current bias conditions because $R_d$ at $T>2$ K is 100 times less than either the cable or free space impedance. The microwave power is kept low enough that $\Delta V$ is linear in power, *assuring that the microwaves only probe existing supercurrents*. The measurement wires are low pass filtered both at room temperature and the sample temperature.

Figure 2(a) shows a family of $\Delta V$ versus $V_{dc}$ curves, at fixed frequency but different $T$. Fluctuations in the microwave power are removed by normalizing $\Delta V$ to a reference detector signal. Clearly, the integer steps have a much stronger $T$-dependence than the half-integer steps. The product of the step width (difference in $I_{dc}$ for the peak and dip of the step) and the step height (difference in $\Delta V$ for the peak and dip), is proportional to $R_d \Lambda^2 (I_{ac})^2$ from which we calculate $\Lambda I_{ac}$. Because $I_{ac}$ remains unknown, we plot $\Lambda$ in arbitrary units [Fig. 2(b)]. The integer step magnitude $\Lambda_1(T)$ has the same exponential $T$-dependence as $I_c(T)$, at least where $I_{dc}>2I_c$ [Fig. 2(b)]. This is consistent with the RSJ model [15]. The much weaker $T$-dependence of the half steps is not contained within the RSJ model, even if generalized to include a nonsinusoidal current-phase relationship. In the RSJ model any ac Josephson effect requires $I_c>0$, [14-15]. In contrast, we observe no vestige of $I_c$ above 5.8 K but resolve the half-integer Shapiro step up to 8.4 K, within 300 mK of the transition temperature of the Nb electrodes, where $2\pi kT=13E_c$.

The integer and half-integer Shapiro steps have very different ν-dependences. Figure 3 shows a series of irradiated response curves at 4.2 K but different ν's. Because $I_{ac}$ is not constant between curves, each curve in

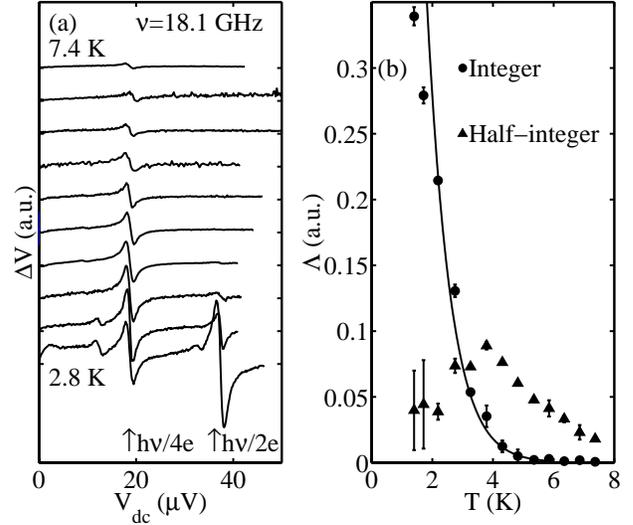

FIG. 2. (a) Normalized $\Delta V$ versus $V_{dc}$ for $T$= 2.8, 3.3, 3.8, 4.3, 4.8, 5.4, 5.9, 6.4, 6.9, 7.4 K. (b) $\Lambda_1$, ●, and $\Lambda_{1/2}$, ▲, versus $T$. Line: $MT^{1/2}\exp[-2\pi T/(4.1 \text{ K})]$ scaled by $M$ to fit $\Lambda_1$, excluding two lowest $T$ points. For linearity in power, the microwave power is reduced ten times for the lowest two $T$ points.

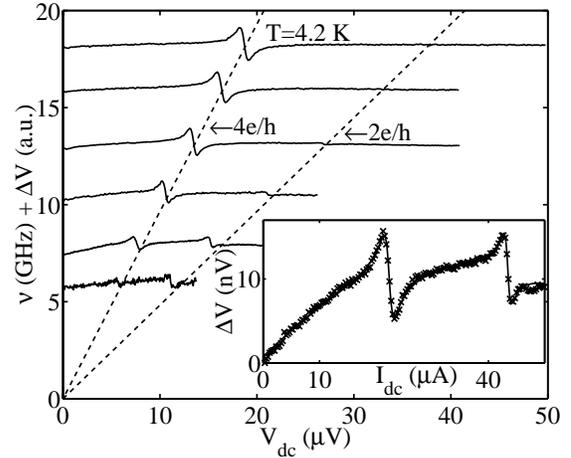

FIG. 3. $\Delta V$ versus $V_{dc}$ for ν=5.4, 7.4, 10.2, 12.9, 15.8, 18.1 GHz. Each curve is normalized to its own maximum value and shifted by ν in GHz. Dotted lines are the ac Josephson relation $\nu=2eV_{dc}/h$ and twice that value $\nu=4eV_{dc}/h$. A background signal causes the apparent deviation of small steps from the ac Josephson relation (see inset). Inset: Unnormalized $\Delta V$ at 7.4 GHz versus $I_{dc}$, **x**, and fit to Eq. 1 plus quadratic background, line.

Fig. 3 only represents the relative magnitude of the integer and half-integer Shapiro steps at a given ν; however, the half-integer steps cannot be seen below 5 GHz while the integer steps are not detectable above 15 GHz. The integer steps shrink with increasing ν ($V_{dc}$), consistent with the RSJ model [15], while the half steps grow.

To extract the magnitude of the half-integer steps $\Lambda_{1/2}$ from these data requires knowledge of $I_{ac}$. Because the behavior of the integer steps is consistent with the RSJ model, we assume that their magnitude is given by the RSJ model and use them to calibrate $I_{ac}$. Fitting Eq. 1 to the data in Fig. 3, but plotted versus $I_{dc}$, we find $R_d\Lambda^2(I_{ac})^2$ for both the integer and half-integer steps (Fig. 3 inset). Assuming $\Lambda_1=I_c/2I_0$ and measuring $R_d$ and $I_c$=1.7 μA from dc data we extract $I_{ac}$ and solve for $\Lambda_{1/2}$. The data are linear in ν ($V_{dc}$), with a slope of (1.8±0.2)×$10^{-3}$ GHz$^{-1}$ [Fig. 4(a)].

In qualitative agreement with the half steps that we observe, theories of NEQ superconductivity in both the ballistic and diffusive limits predict NEQ ac supercurrents that flow at multiples of the ac Josephson frequency, grow with increasing $V_{dc}$ and are only algebraically suppressed with $T$ in contrast to the equilibrium supercurrents which are exponentially suppressed [5-6]. Accompanying these NEQ ac supercurrents is an enhancement of the device conductance around $V_{dc}$=0, [5-7] which we also observe (Fig. 4).

These qualitative features may be understood in terms of simple physical arguments. Equilibrium supercurrents are suppressed exponentially when $kT>>E_c$, because the population of initially phase coherent quasiparticles have an energy distribution in the normal metal a few $kT$ wide. Their ensemble phase coherence is lost in a characteristic time $\hbar/kT$ while a typical quasiparticle requires $\hbar/E_c=\tau_d$ to cross the junction. This thermal smearing may be overcome by introducing NEQ structure in the electron distribution function with a characteristic energy scale $E_c$, [7].

NEQ populations in the normal metal arise because the energy $E_n$ ($E_{-n}$) of the $n$th quasiparticle state, or Andreev bound (AB) state, above (below) the Fermi energy, $E_F\equiv0$, is a periodic function of $\phi$, with typical amplitude $E_c$ or less.[5,16-17]. The ac Josephson relation, $d\phi/dt=2eV/\hbar$, implies $E_n(\phi)$ and its equilibrium Fermi-Dirac population $f_{eq}(E_n)$ oscillate in time at a frequency $2eV_{dc}/\hbar$. A relaxation-time approximation, $df/dt=-(1/\tau_E)(f-f_{eq})$, models the instantaneous occupation $f$ lagging $f_{eq}$ by the energy relaxation-time $\tau_E$. For $V_{dc}<\hbar/2\tau_Ee$, the deviation from equilibrium and any NEQ supercurrents grow with increasing $V_{dc}$ [5,7,18], [Fig. 4(a)].

The phase dependence of the spectrum of AB states has structure on energy scales comparable to $E_c$. In particular, the AB spectrum has an $E_c$ wide "energy gap" about $E_F$ when $\phi=0$ which closes when $\phi=\pi$, [5,7,16-17]. This imposes features on the instantaneous deviation from equilibrium $f(E_n)-f_{eq}(E_n)$ which are about $E_c$ wide; therefore, the ensemble phase coherence for this NEQ portion of the AB state population will decay with the characteristic time $\hbar/E_c=\tau_d$ and the NEQ supercurrent will not be exponentially suppressed with increasing $T$, [5-7].

The NEQ supercurrent will have an explicit $T$ dependence through the amplitude of $f_{eq}(E_n(\phi))$. When $kT>>E_c$ then $f_{eq}\approx1/2-E_n(\phi)/4kT$ and when $kT<<E_c$ then $f_{eq}\approx\exp(-E_n(\phi)/kT)$ implying the $f_{eq}(\phi)$ are maximally modulated when $kT\approx E_c$ and have a $1/T$ suppression at high $T$. The half-integer steps we measure are suppressed at both low and high temperatures with a maximum around 4 K [Fig. 2(b)] which is consistent with our estimate of $E_c$=4.1 K, though the analysis ignores the implicit $T$ dependence in $\tau_E$ and the complication of large equilibrium supercurrents at low $T$.

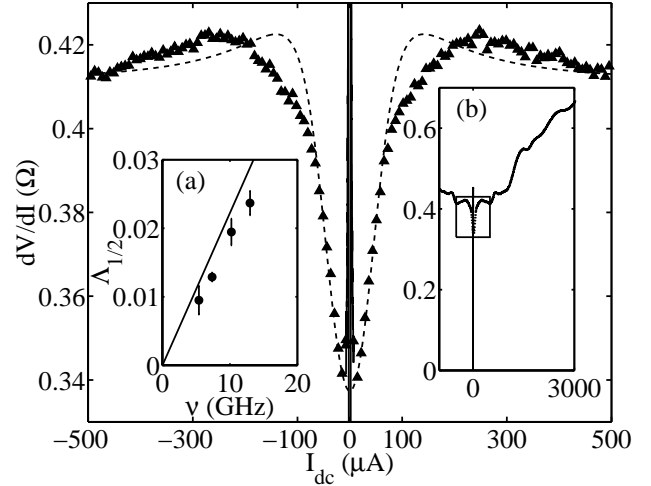

FIG. 4. $dV/dI$ at 4.2 K versus $I_{dc}$, ▲ and solid line, and the NEQ theory fit, dashed line. Data points suppressed in central region. Abscissa extended to 500 μA ($V_{dc}\approx200$ μV) to show convergence of fit to data where $R_N=dV/dI$. Insets: (a) $\Lambda_{1/2}$ measured at 4.2 K, ●, and predicted by the NEQ model, line. (b) $dV/dI$ on a larger $I_{dc}$ scale, boxed area shows region of main figure.

The supercurrent carried by a single AB state is $(2e/\hbar)(dE_n/d\phi)f$ [5,16] which can be argued by equating the energy spent by an external source $IVdt$ to the energy stored in the AB states $\Sigma_n dE_n f(E_n)$. Explicitly separating equilibrium and NEQ components and suppressing the $n$-dependence of $f$ and $f_{eq}$, the total supercurrent $I_s$ is $(2e/\hbar)\Sigma_n(dE_n/d\phi)[f_{eq}+(f-f_{eq})]$. The NEQ term is a product of the two $\phi$-periodic factors, $dE_n/d\phi$ and $(f-f_{eq})$, generating strongly $2\phi$-periodic ac supercurrents which flow at twice the ac Josephson frequency and can be detected as half-integer Shapiro steps. For $V_{dc}<<\hbar/2\tau_Ee$, $(f-f_{eq})\propto(dE_n/d\phi)(d\phi/dt)$ which produces a significant dc component in the NEQ part of $I_s$, observed as a conductance enhancement around $V_{dc}$=0, (Fig. 4), [3,5,7]. The equilibrium term will also generate fractional steps but they will be exponentially suppressed both at $T>>E_c$, due to thermal smearing, and at $T<<E_c$, through the $T$ dependence of $f_{eq}$. This may account for the small 1/3 step that we resolve between 2.3 K and 3.8 K [Fig. 2(a)].

Because our junction is neither in an ideal diffusive nor ideal ballistic limit, the spectrum of AB states in our junction will differ from those calculated in Refs. [5,16-17]. As suggested in Ref. 5, we may use an RSJ-like model and general properties of AB states to quantitatively predict the half-integer Shapiro steps from the zero-bias conductance enhancement. We take $E_n=A_n\cos\phi+B_n$ for each $n$, with $E_{-n}=-E_n$ [5,16-17], and ignore higher harmonics of $\phi$ for simplicity, even though they could model the very weak 1/4 step visible in Fig. 2(a). In the high $T$ limit, $kT \gg E_c$, all states with an appreciable $A_n$ have $f_{eq} \approx 1/2 - E_n/4kT$, [5,17] and therefore $f \approx 1/2 - (A_n F + B_n)/4kT$, where $F$ is defined by $dF/dt = (1/\tau_E)(\cos\phi - F)$. Our RSJ-like NEQ model becomes

$$\frac{d\phi}{dt} = \frac{2eR_N}{\hbar}(I - I_s), I_s = -2I_{neq}\sin\phi(\cos\phi - F); \qquad (2)$$

where $I_{neq} = (e/\hbar)\Sigma_n(A_n)^2/4kT$ is treated as an additional phenomenological parameter because we have no experimental access to the individual $A_n$, and $R_N$ is the resistance in the absence of coherent effects. In order to accurately track $f$, a low voltage limit, $eV_{dc} \ll (E_c\hbar/\tau_E)^{1/2}$, is implicit in such a model [5].

The NEQ model can be integrated numerically to give $\Lambda_{1/2}(\nu)$ and the $dV/dI$. We compare $\Lambda_{1/2}(\nu)$ observed to $\Lambda_{1/2}(\nu)$ predicted by the NEQ model, with the model parameters $R_N$, $I_{neq}$, and $\tau_E$, extracted from the $dV/dI$. The theoretical $dV/dI$ which shows enhanced conductance at zero bias $G_0 = G_N + \tau_E(2e/\hbar)I_{neq}$, where $G_N = 1/R_N$ [5], cf. [3,6-7], can be fit to the data at 4.2 K. The data points within 5 μA of zero bias are ignored as they are influenced by the vestigial $I_c$ (Fig. 4). We fit the data only in the region of $|V_{dc}| < 50$ μV because the theory is valid for $eV_{dc} \ll (E_c\hbar/\tau_E)^{1/2} \approx 150$ μeV [5]. From this fit $\tau_E = 10$ ps and $I_{neq} = 17$ μA. Because of the close correspondence between the CVC and the RSJ model [Fig. 1(a)] over a range of $|V_{dc}| < 200$ μV we choose $R_N = 0.411$ Ω, the differential resistance at $V_{dc} = 200$ μV, below the onset of a feature in the $dV/dI$ centered at $|V_{dc}| \approx E_c/e$ (Fig. 4).

With these parameters we integrate the NEQ model with a weak applied ac current, generating Shapiro steps and determining $\Lambda_{1/2}(\nu)$ theoretically. In the region studied it is linear with a slope $(2.2 \pm 0.2) \times 10^{-3}$ GHz$^{-1}$. The good agreement with the measurement is shown [Fig. 4(a)]. We also resolve four SGS peaks [Fig. 1(c)], providing the estimate $\tau_E \approx 4\tau_d = 7$ ps [19], in reasonable agreement with the model. This good agreement does not preclude additional effects that might simply be absorbed into the model's phenomenological parameters.

In summary, we have observed that in clean mesoscopic SNS junctions, an ac supercurrent persists when $I_c = 0$. This ac supercurrent flows at twice the Josephson frequency, and is in qualitative agreement with NEQ theories. A NEQ model, with parameters extracted from the zero bias conductance enhancement, correctly predicts the magnitude of the NEQ ac supercurrent.

We thank J.G.E. Harris and T.M. Klapwijk for useful discussions. This work is supported by the NSF through the NSF Science and Technology Center for Quantized Electronic Structures, Grant No. DMR-91-20007, and the ONR, grant No. N0014-92-J-1452.